\documentclass[
]{ceurart}

\usepackage{listings}
\begin{document}

\copyrightyear{2022}
\copyrightclause{Copyright for this paper by its authors.
  Use permitted under Creative Commons License Attribution 4.0
  International (CC BY 4.0).}

\conference{Woodstock'22: Symposium on the irreproducible science,
  June 07--11, 2022, Woodstock, NY}

\title{Multilingual Prompts in LLM-Based Recommenders: Performance Across Languages}





\author[1]{Makbule Gulcin Ozsoy}[%
email=makbulegulcin@gmail.com,
]
\fnmark[1]
\address[1]{London, UK}


\begin{abstract}
Large language models (LLMs) are increasingly used in natural language processing tasks. Recommender systems traditionally use methods such as collaborative filtering and matrix factorization, as well as advanced techniques like deep learning and reinforcement learning. Although language models have been applied in recommendation, the recent trend have focused on leveraging the generative capabilities of LLMs for more personalized suggestions. While current research focuses on English due to its resource richness, this work explores the impact of non-English prompts on recommendation performance. Using OpenP5, a platform for developing and evaluating LLM-based recommendations, we expanded its English prompt templates to include Spanish and Turkish. Evaluation on three real-world datasets, namely ML1M, LastFM, and Amazon-Beauty, showed that usage of non-English prompts generally reduce performance, especially in less-resourced languages like Turkish. We also retrained an LLM-based recommender model with multilingual prompts to analyze performance variations. Retraining with multilingual prompts resulted in more balanced performance across languages, but slightly reduced English performance. This work highlights the need for diverse language support in LLM-based recommenders and suggests future research on creating evaluation datasets, using newer models and additional languages.
\end{abstract}

\begin{keywords}
  Recommender systems \sep
  Large language models \sep
  Prompting \sep
  Multi-language evaluation
\end{keywords}

\maketitle

\section{Introduction} \label{intro}


Large language models (LLMs) have become essential tools in natural language processing, excelling in tasks like named entity recognition, text classification, summarisation, and translation. 
They are designed to understand and generate natural language, making it more intuitive for end-users to interact with machines through applications like search engines and chatbots. Recently, there is increasing interest in using LLMs within recommender systems.
Recommender systems estimate user preferences to suggest relevant items, using various techniques, such as collaborative filtering, content-based filtering, and matrix factorization \cite{pan2008one, Ye2010, li2015rank, HeLLSC16}, deep learning \cite{ozsoy2016word, vasile2016meta, he2017neural, musto2018deep}, reinforcement learning \cite{zheng2018drn, shi2021dares}, and language models-based \cite{sun2019bert4rec, fan2023recommender} methods. 
The recent trend is to leverage LLMs' generative capabilities for more personalized suggestions. 
Researchers are investigating how LLMs can enhance recommendations by understanding complex user behaviors and language patterns.

\begin{figure}
    \centering
    \includegraphics[width=0.8\linewidth]{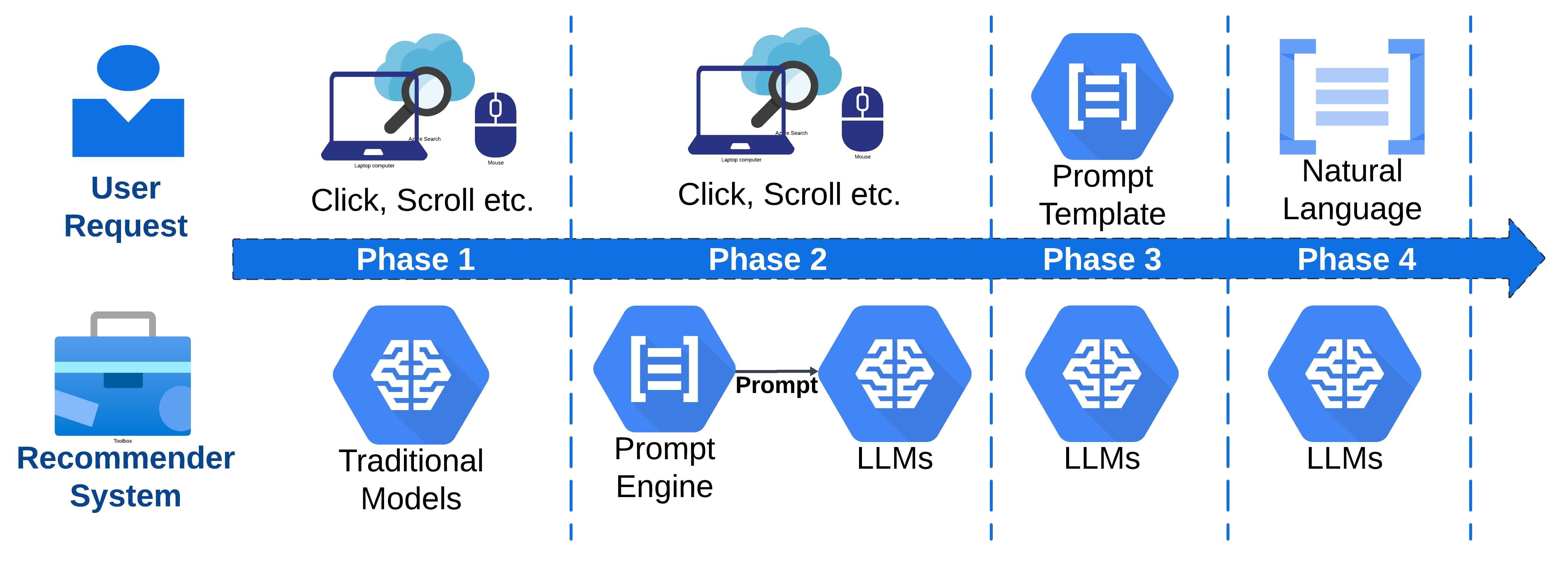}
    \caption{Four phases in the evolution of LLM-based recommenders: (i) Initial Phase: Traditional methods generate suggestions based on user actions like clicks. (ii) LLM Integration: LLMs enhance recommendations using prompts. (iii) Prompt Template Interaction: Users request recommendations via natural language templates. (iv) Natural Language Interaction: Direct native language communication with the system.}
    \label{fig:phases}
\end{figure}

Given this trend, we anticipate four phases in the evolution of LLM-based recommenders, as shown in Figure \ref{fig:phases}: 
(i) Initial Phase: Users request recommendations through actions like clicking, and black-box recommender systems use traditional methods to generate personalized suggestions.
(ii) LLM Integration: In this phase, LLMs are introduced into the recommender system. Although users continue to interact via clicks or scrolls, the system now uses prompts and LLM models to generate recommendations. Prompts are in the language where the LLM performs best, such as English.
(iii) Prompt Template Interaction: Users interact with the system using natural language through prompt templates. Instead of relying only on clicks or scrolls, users fill out templates to request recommendations. While in the initial stages, these prompts could be in English, later they could be in the user's native language.
(iv) Natural Language Interaction: Users interact with the recommender system directly in their native language, without the need for prompt templates. 
Considering recent developments in LLM-based recommenders, such as those in \cite{geng2022recommendation, xu2024openp5, ngo2024recgpt}, it appears the field is already in phase two or even phase three. 
However, these phases often rely on user interactions in languages where LLMs perform best, primarily English. Current state-of-the-art LLMs cover only a small percentage of the world's spoken languages, favoring those with abundant resources like English \cite{nicholas2023lost, zhao2024llama, Cohere24AILangGap}.

This work investigates how non-English prompts affect recommendation performance using the OpenP5 platform \cite{xu2023openp5, xu2024openp5}, which supports developing, training, and evaluating LLM-based models for generative recommendation. We expanded OpenP5's English templates to include Spanish and Turkish, assessing their impact on performance. Spanish, with more resources than Turkish but fewer than English, was expected to perform slightly worse, while Turkish, being less resourced, was anticipated to perform lower.
In addition to using non-English prompts on a pretrained model, we examined the effects of further training the model with multilingual prompts, namely in English, Spanish, and Turkish. 
The main contributions of this work are: 
\begin{itemize}
    \item Exploring the impact of non-English prompts on LLM-based recommenders by comparing performance with English, Spanish, and Turkish prompts.
    \item Exploring the effects of further training an LLM-based recommender with multilingual prompts. 
    \item Evaluating LLM-based recommenders on three real-world datasets, namely ML1M, LastFM and Amazon-Beauty, finding that non-English prompts generally reduce performance particularly for languages less similar to English. However, retraining the model with multiple languages led to a more balanced performance, with a slight decrease for English prompts. 
\end{itemize}

The paper is organized as follows: Section \ref{relWork} provides background on recommender models, with a focus on LLM-based systems, their challenges, and LLMs on non-English content. Section \ref{llm_based_rec_beyond_en} explores the effects of non-English prompts on LLM-based recommendation systems, including (i) the use of non-English prompts on an already trained LLM-based recommender and (ii) the impact of training a new model with both English and non-English prompts. Section \ref{discuss} analyzes and discusses the experimental results, while Section \ref{conclusion} presents conclusion and outlines future research directions.

\section{Related Work}\label{relWork} 

This section provides an overview of LLM-based recommender systems, the challenges they encounter, and performance of LLMs with non-English content. 

\subsection{Integration of Recommender Systems and LLMs}
Recommender systems aim to estimate users' preferences and recommend items based on historical user-item interaction data. They can use various approaches, such as traditional collaborative filtering, content-based filtering, matrix factorization \cite{pan2008one, Ye2010, li2015rank, HeLLSC16}, deep learning \cite{ozsoy2016word, vasile2016meta, he2017neural, musto2018deep}, reinforcement learning \cite{zheng2018drn, shi2021dares}, and language model-based methods \cite{sun2019bert4rec, qiu2021u, zhang2022gbert, fan2023recommender, li2023text}. 
Recently, researchers have started to integrate capabilities of LLMs in recommender systems\cite{fan2023recommender, li2024large, vats2024exploring}. 

\paragraph{LLM-Based Recommenders and Their Challenges}
For integrating recommender systems and LLMs, researchers utilize two key strategies: non-tuning-based methods and tuning-based methods. 
Non-tuning-based methods \cite{dai2023uncovering, sun2023chatgpt, gao2023chat, he2023large, zhang2023chatgpt, wang2023zero, du2024enhancing, liu2024once, hou2024large} utilize in-context learning and prompt optimization to interact with LLMs (e.g., Chat-GPT1) without training. The main focus in these works is the design of prompts, which require extensive expert knowledge and human labor. Additionally, they might not perform as well as traditional recommendation methods \cite{liu2023chatgpt, tan2024idgenrec}.
Tuning-based methods focus on further training LLMs with domain-specific knowledge. These approaches leverage historical interactions and contextual data, incorporating prompts and instructions during the tuning process, though they may suffer from high computational costs. For example,  P5 \cite{geng2022recommendation}, RecSysLLM\cite{chu2023leveraging}, TALLRec \cite{bao2023tallrec}, PALR\cite{yang2023palr}, PBNR\cite{li2023pbnr}, InstructRec\cite{zhang2023recommendation}, GenRec \cite{ji2024genrec}, RecGPT\cite{ngo2024recgpt}, Re2LLM \cite{wang2024re2llm}, GLRec \cite{wu2024exploring} use various prompts to fine-tune LLMs for recommendation purposes. Additionally, PPR \cite{wu2024personalized} and UniCRS\cite{wang2022towards} have explored prompt learning techniques. 
These approaches leverage historical interactions and contextual data and incorporate prompts and instructions during the tuning process. However, they may suffer from high computational costs.
Other research efforts aim to support researchers working with LLM-based recommenders. For example, OpenP5 \cite{xu2023openp5, xu2024openp5} has developed a platform for exploring LLM-based recommendations, providing an environment for the development, training, and evaluation of generative recommender systems for research purposes.

\paragraph{Challenges of LLM-based Recommenders}
The approaches explained above focus on improving the recommendations performance. However, LLM-based recommenders have some additional challenges. 
Position bias occurs since LLMs favor items at the top of the list 
\cite{wu2023survey}. Popularity bias is another issue, where frequently mentioned items in training data tend to rank higher, potentially reducing diversity and worsening cold-start problems \cite{wu2023survey}. Fairness bias is also a concern, as LLMs may exhibit biases related to sensitive attributes such as age, gender, or race \cite{wu2023survey, li2024large}. 
Additionally, LLMs face challenges in safety and robustness, as small changes can affect their reliability, risking manipulation or misuse\cite{fan2023recommender}. Hallucinations are another problem, where LLMs might recommend non-existent items, leading to user dissatisfaction \cite{azamfirei2023large, li2024large}. Effective prompt design is crucial, as converting user and item attributes into natural language prompts can be limited by context length. 
Controlling LLM outputs to meet specific constraints (e.g., price, color) is difficult, and ensuring consistent formatting is challenging \cite{wu2023survey, tan2023user}. Finally, privacy concerns arise because LLMs use large datasets that may contain sensitive user information, risking data exposure \cite{fan2023recommender}.
Even though various studies have explored many challenges in LLM-based recommenders \cite{fan2023recommender, li2024large, vats2024exploring, wu2023survey}, the impact of prompt language has not been thoroughly investigated. Most LLM-based recommender systems predominantly use English prompts, whether for tuning or non-tuning methods. Some notable exceptions include RecSysLLM \cite{chu2023leveraging}, which uses English and Chinese prompts; both high-resource languages \cite{nicholas2023lost}; and FaiRLLM \cite{zhang2023chatgpt}, which examines user country information but uses a consistent prompt language throughout. In this work, we investigate how prompt language affects recommendation performance by exploring languages that are relatively low-resource and linguistically distinct from English.

\subsection{LLMs and Non-English Content}
Large language models (LLMs) have become the dominant tools for performing various natural language processing tasks. Even though they are mainly build on English texts, researchers extend their capabilities to other non-English languages by building multi-language models \cite{nicholas2023lost}. These models are trained on multiple languages simultaneously and can infer connections between them. 
As the result, they can utilize word associations and grammatical rules acquired from languages with more resources, such as English, and apply them to languages with less available data. 
However, these models are predominantly trained on English text, leading them to transfer values and assumptions encoded in English into other language contexts \cite{nicholas2023lost, zhao2024llama, Cohere24AILangGap}.
Not all languages have the same resourcedness levels, such that the volume, quality, and diversity of the available data to train language models are different. 
The disparity in resourcedness levels means that, even when trained on multiple languages, LLMs perform significantly better in higher-resource languages and those similar to them compared to lower-resource languages \cite{nicholas2023lost}. 
There are research works that are focusing on challenges in low-resource languages. 
Some researchers collect data in specific languages and retrain or fine-tune LLMs \cite{team2023internlm, cui2023efficient, acikgoz2024bridging, kesgin2024introducing}. Zhao et al. \cite{zhao2024llama} further analyzed how expanding vocabulary, additional pre-training, and instruction tuning affect the ability of LLMs to generate text and follow instructions in a non-English language. They showed that comparable performance can be achieved with less than 1\% of the pre-training data. 


\section{LLM-based Recommenders Beyond English} \label{llm_based_rec_beyond_en}
We explore the impact of non-English prompts on LLM-based recommender systems in two ways: (i) applying non-English prompts to an existing model and (ii) using both English and non-English prompts to further train the model. 
In this section, we will first describe the evaluation setup and then discuss the tasks and their results.

\subsection{Experimental Setup}

\begin{table}
\caption{The statistics of the datasets}\label{table:dataset}
\begin{center}
\begin{tabular}{ccccc} 
 \toprule
 {} & \textbf{\#Users} & \textbf{\#Items} & \textbf{\#Interactions} & \textbf{Sparsity} \\
 \midrule
 \textbf{ML1M} & 6,040 & 3,616 & 999,611 & 0.9516 \\
 \textbf{LastFM} & 1,090 & 3,646 & 52,551  & 0.9868\\
 \textbf{Amazon Beauty} & 22,363 & 12,101 & 198,502 & 0.9993 \\
\bottomrule
 \end{tabular}
\end{center}
\end{table}

We used OpenP5 \cite{xu2023openp5, xu2024openp5} platform, which is recently developed for developing, training, and evaluating LLM-based models for generative recommendation. 
It contains implementations of an encoder-decoder LLM, namely T5 \cite{raffel2020exploring}, and a decoder-only LLM, namely Llama-2 \cite{touvron2023llama} and provides 10 widely recognized public datasets for experiments. 
For indexing the items, it utilizes three item indexing methods: (i) Random indexing where the item ids are assigned randomly. It is fairly simple method but can introduce artificial relationships among items due to the way tokenizers used by LLMs work. (ii) Sequential indexing which assigns item ids to consecutive numbers based on users’ consecutive interactions observed in training data.  This approach mitigates tokenizer issues, as items with the same prefix are likely related, reflecting actual user interactions. (iii) Collaborative indexing which assigns item ids by the frequency of co-occurrence of items.
For recommendation, OpenP5 focuses on sequential and straightforward recommendation tasks, where the former utilizes user ids along with user history, while the latter relies only on user ids.
For each recommendation task, it offers eleven personalized prompts templates, all of which are created by the authors and are all in English. During execution, whether for training or making recommendations, these prompts are filled with personalized information for each user, such as user ids and interaction history as a list of item ids.
The prompts are divided into seen and unseen categories. Out of the eleven prompts, one is designated as the unseen prompt and is used to assess the model's zero-shot generalization capability.
In addition to offering a platform for LLM-based recommendations, OpenP5 also provides evaluation results. In their experiments, they fine-tuned the T5 \cite{raffel2020exploring} model on full parameters, and Llama-2 \cite{touvron2023llama} using the LORA \cite{hu2021lora} technique.

In this work, we used random (R) and sequential (S) indexing techniques, omitting collaborative indexing as we couldn't achieve consistent performance, even with English prompts. We employed the OpenP5-T5 model, a fine-tuned version of T5 \cite{raffel2020exploring} optimized for recommendation tasks.
Experiments were conducted on three well-known recommendation datasets, namely ML1M, LastFM, and Amazon Beauty, which are pre-split, pre-processed and publicly available \footnote{\url{https://github.com/agiresearch/OpenP5}}. 
The statistics of the datasets are summarized in Table~\ref{table:dataset}. 
Performance was measured using HitRate and NDCG metrics @5 and @20, following the OpenP5 framework \cite{xu2023openp5, xu2024openp5}.
Computations were performed on a P100 GPU with a 30-hour weekly usage limit. All experiments were re-run, and results are discussed in the following sections.

\begin{table}
\caption{Command asking translations of all prompts, and an example translation} \label{table:prompt_translate}
\begin{center}
\begin{tabular}{p{7em}p{28.5em}} 
 \toprule
 \multirow{2}{*}{\parbox{6em}{\textbf{Translation}\\\textbf{command}}} & Translate following sentences to Spanish: """1. Considering \{dataset\} user\_\{user\_id\} has interacted with \{dataset\} items \{history\} . What is the next recommendation for the user ?; \{dataset\} \{target\} ...""" \\
 \multirow{2}{*}{\parbox{6em}{\textbf{Example}\\\textbf{translation}}}  &  Considerando que el user\_\{user\_id\} de \{dataset\} ha interactuado con \{history\} elementos de \{dataset\}. ¿Cuál es la próxima recomendación para el usuario?; \{dataset\} \{target\} \\
 
\bottomrule
 \end{tabular}
\end{center}
\end{table}

\begin{table}
\caption{Performance on ML1M dataset with [Random (R), Sequential (S)] indexing, [seen (s), unseen (u)] prompts for [Sequential, Straightforward] recommendation in [EN: English, ES: Spanish, TR: Turkish] (H: Hitrate, N: NDCG)}\label{table:compare_prompts_ml1m}
\begin{center}
\begin{tabular}{c|c|cccc|c|cccc|c} 
 \toprule
 \multirow{2}{*}{\parbox{3.5em}{\centering Indexing\\(Prompt)}} & 
 \multirow{2}{*}{Lng} & 
        \multicolumn{5}{c}{Sequential} & \multicolumn{5}{c}{Straightforward} \\ 
        \cline{3-7}  \cline{8-12} 
{} & {}     & H@5 & N@5 & H@10 & N@10 & Avg\% & H@5 & N@5 & H@10 & N@10 & Avg\%\\
\midrule 
\multirow{3}{*}{R (s)} & EN & 0.1098 & 0.0734 & 0.1573 & 0.0888 & - & 0.0215 & 0.0133 & 0.0346 & 0.0175 &  - \\
{} & ES & 0.1118 & 0.0728 & 0.1586 & 0.0879 & 0.21& 0.0220 & 0.0140 & 0.0351 & 0.0181  & 3.12\\
{} & TR & 0.1071 & 0.0722 & 0.1543 & 0.0875 & -1.87& 0.0207 & 0.0128 & 0.0343 & 0.0171 & -2.66\\
\midrule 
\multirow{3}{*}{R (u)}& EN & 0.1060 & 0.0693 & 0.1533 & 0.0846 & - & 0.0219 & 0.0138 & 0.0341 & 0.0177  & - \\
{} & ES & 0.1061 & 0.0700 & 0.1561 & 0.0862 & 1.21 & 0.0219 & 0.0136 & 0.0364 & 0.0183 & 2.17\\
{} & TR & 0.0232 & 0.0148 & 0.0364 & 0.0190  & -77.64 & 0.0220 & 0.0139 & 0.0354 & 0.0182 & 1.95\\
\midrule 
\midrule 
\multirow{3}{*}{S (s)} & EN & 0.2101 & 0.1430 & 0.3053 & 0.1737 & - & 0.0310 & 0.0192 & 0.0571 & 0.0275 & -\\
{} & ES & 0.1490 & 0.0982 & 0.2220 & 0.1218 & -29.39& 0.0320   & 0.0198 & 0.0568 & 0.0277 & 1.64\\
{} & TR & 0.0624 & 0.0399 & 0.0985 & 0.0515& -70.12  & 0.0286  & 0.0183 & 0.0561 & 0.0272 & -3.82\\
\midrule 
\multirow{3}{*}{S (u)} & EN & 0.2116 & 0.1436 & 0.3055 & 0.1737 & - & 0.0316 & 0.0193 & 0.0566 & 0.0272  & -\\
{}& ES & 0.1697 & 0.1137 & 0.2493 & 0.1394 & -19.69  & 0.0311 & 0.0193 & 0.0579 & 0.0279 & 0.82\\
{} & TR & 0.0296 & 0.0191 & 0.0517 & 0.0261 & -85.19  & 0.0298 & 0.0186 & 0.0556 & 0.0269 & -3.05\\
\bottomrule
 \end{tabular}
\end{center}
\end{table}

\begin{table}
\caption{Performance on LastFM dataset with [Random (R), Sequential (S)] indexing, [seen (s), unseen (u)] prompts for [Sequential, Straightforward] recommendation in [EN: English, ES: Spanish, TR: Turkish] (H: Hitrate, N: NDCG)}\label{table:compare_prompts_lastfm}
\begin{center}
\begin{tabular}{c|c|cccc|c|cccc|c} 
 \toprule
\multirow{2}{*}{\parbox{3.5em}{\centering Indexing\\(Prompt)}} & 
 \multirow{2}{*}{Lng} & 
        \multicolumn{5}{c}{Sequential} & \multicolumn{5}{c}{Straightforward} \\ 
        \cline{3-7}  \cline{8-12} 
{} & {}     & H@5 & N@5 & H@10 & N@10 & Avg\% & H@5 & N@5 & H@10 & N@10 & Avg\%\\
\midrule 
\multirow{3}{*}{R (s)} & EN & 0.0156 & 0.0104 & 0.0312 & 0.0153 & - & 0.0239 & 0.0151 & 0.0294 & 0.0169 & -\\
{} & ES & 0.0165 & 0.0094 & 0.0303 & 0.0139 & -3.97 & 0.0202 & 0.0121 & 0.0339 & 0.0167  & -5.31\\
{} & TR & 0.0174 & 0.0118 & 0.0266 & 0.0147 & 1.59  & 0.0202 & 0.0128 & 0.0312 & 0.0164 & -6.89\\
\midrule 
\multirow{3}{*}{R (u)}& EN & 0.0128 & 0.0072 & 0.0248 & 0.0110 & -  & 0.0211 & 0.0130 & 0.0358 & 0.0178 & -\\
{} & ES & 0.0147 & 0.0087 & 0.0294 & 0.0135 & 19.24  & 0.0229 & 0.0137 & 0.0349 & 0.0176 & 2.57\\
{} & TR & 0.0184 & 0.0121 & 0.0312 & 0.0163 & 46.45 & 0.0220 & 0.0134 & 0.0294 & 0.0158  & -5.44\\
\midrule 
\midrule 
\multirow{3}{*}{S (s)} & EN & 0.0395 & 0.0262 & 0.0587 & 0.0323 & - & 0.0376 & 0.0259 & 0.0661 & 0.0350 & -\\
{} & ES & 0.0321 & 0.0215 & 0.0523 & 0.0280 & -15.22  & 0.0330 & 0.0212 & 0.0578 & 0.0293 & -14.81\\
{} & TR & 0.0275 & 0.0167 & 0.0459 & 0.0226 & -29.62  & 0.0312 & 0.0221 & 0.0523 & 0.0289 & -17.50\\
\midrule 
\multirow{3}{*}{S (u)} & EN & 0.0403 & 0.0265 & 0.0606 & 0.0331 & - & 0.0403 & 0.0282 & 0.0679 & 0.0370 & - \\
{}& ES & 0.0358 & 0.0230 & 0.0532 & 0.0288 & -12.40& 0.0376 & 0.0239 & 0.0551 & 0.0297  & -15.13\\
{} & TR & 0.0275 & 0.0195 & 0.0468 & 0.0256 & -25.90  & 0.0330 & 0.0212 & 0.0505 & 0.0269 & -23.97\\
\bottomrule
 \end{tabular}
\end{center}
\end{table}

\begin{table}
\caption{Performance on Amazon-Beauty dataset with [Random (R), Sequential (S)] indexing, [seen (s), unseen (u)] prompts for [Sequential, Straightforward] recommendation in [EN: English, ES: Spanish, TR: Turkish] (H: Hitrate, N: NDCG)}\label{table:compare_prompts_beauty}
\begin{center}
\begin{tabular}{c|c|cccc|c|cccc|c} 
 \toprule
\multirow{2}{*}{\parbox{3.5em}{\centering Indexing\\(Prompt)}} & 
 \multirow{2}{*}{Lng} & 
        \multicolumn{5}{c}{Sequential} & \multicolumn{5}{c}{Straightforward} \\ 
        \cline{3-7}  \cline{8-12} 
{} & {}     & H@5 & N@5 & H@10 & N@10 & Avg\% & H@5 & N@5 & H@10 & N@10 & Avg\%\\
\midrule 
\multirow{3}{*}{R (s)} & EN & 0.0318 & 0.0226 & 0.0463 & 0.0273 & - & 0.0231 & 0.0168 & 0.0316 & 0.0195 & -\\
{} & ES & 0.0288 & 0.0204 & 0.0427 & 0.0249 & -8.93& 0.0190 & 0.0141 & 0.0284 & 0.0172  & -13.94\\
{}& TR & 0.0262 & 0.0178 & 0.0383 & 0.0217 & -19.16& 0.0207 & 0.0151 & 0.0292 & 0.0179  & -9.08\\
\midrule 

\multirow{3}{*}{R (u)} & EN & 0.0314 & 0.0222 & 0.0457 & 0.0268 & - & 0.0222 & 0.0164 & 0.0312 & 0.0193 & -\\
{} & ES & 0.0251 & 0.0167 & 0.0380 & 0.0209 & -20.92& 0.0203 & 0.0150 & 0.0286 & 0.0177  & -8.43\\
{} & TR & 0.0207 & 0.0152 & 0.0289 & 0.0178 & -33.99& 0.0213 & 0.0153 & 0.0297 & 0.0180  & -5.58\\
\midrule 
\midrule 

\multirow{3}{*}{S (s)}& EN & 0.0456 & 0.0335 & 0.0622 & 0.0389 & - & 0.0318 & 0.0239 & 0.0439 & 0.0278 & -\\
{}& ES & 0.0403 & 0.0293 & 0.0563 & 0.0345 & -11.24& 0.0295 & 0.0219 & 0.0409 & 0.0256  & -7.59\\
{} & TR & 0.0369 & 0.0261 & 0.0513 & 0.0307 & -19.94& 0.0301 & 0.0227 & 0.0406 & 0.0261  & -6.00\\
\midrule 
\multirow{3}{*}{S (u)} & EN & 0.0454 & 0.0332 & 0.0614 & 0.0385 & - & 0.0314 & 0.0237 & 0.0436 & 0.0276  & -\\
{} & ES & 0.0379 & 0.0276 & 0.0524 & 0.0322 & -16.10& 0.0267 & 0.0197 & 0.0382 & 0.0234  & -14.87\\
{} & TR & 0.0280 & 0.0215 & 0.0390 & 0.0250 & -36.28  & 0.0300 & 0.0228 & 0.0411 & 0.0263 & -4.68\\
\bottomrule
 \end{tabular}
\end{center}
\end{table}

\subsection{Impact of Non-English Prompts on a PreTrained LLM-Based Recommender}
In order to explore the impact of non-English prompts on a pretrained LLM-based recommender, we used prompts in Spanish and Turkish alongside English. 
English is a extremely-high resource language by several orders of magnitude, while Spanish is a high-resource and Turkish is medium-resource language \cite{nicholas2023lost, joshi2020state}. 
Previous research \cite{dhamecha2021role} shows that linguistic similarity influences model performance. 
Spanish and English belong to the Indo-European language family, whereas Turkish is part of the Altaic language family. Given these language characteristics, we predict that the performance of Spanish prompts will be slightly lower than that of English prompts but still relatively close. In contrast, using Turkish prompts is expected to decrease the performance of the LLM-based recommender systems. 
The Spanish and Turkish prompts were generated using machine translation through the ChatGPT interface with the GPT-3.5 model. We used a command to translate all prompts into Spanish and Turkish. The command used and an example translation are shown in Table \ref{table:prompt_translate}. 
In the experiments, translated prompts were input into a pretrained model, and recommendation performance across different languages was compared.
The performance comparisons on ML1M, LastFM and Amazon-Beauty datasets are presented in Table~\ref{table:compare_prompts_ml1m}, Table~\ref{table:compare_prompts_lastfm} and Table~\ref{table:compare_prompts_beauty}, respectively. The tables show the evaluation metrics results and the average changes across all metrics compared to the English prompts.  

\paragraph{Observations}
When we analyze different prompt types (seen or unseen), recommendation types (sequential or straightforward), and datasets (ML1M, LastFM, Amazon Beauty), we observe that the behavior of random indexing is inconsistent and varies across datasets. 
For example, with seen prompts, the average change in evaluation metrics for Spanish and Turkish prompts remains relatively stable ($\pm5\%$) on the ML1M and LastFM datasets. However, on the Amazon Beauty dataset, these changes are more noticeable, ranging from  $-9\%$ to $-19\%$. 
In contrast, when using unseen prompts, performance with straightforward recommendations remains more stable compared to sequential recommendations across all datasets. 
We attribute this behavior to the characteristics of the random indexing. As discussed in OpenP5 \cite{xu2023openp5, xu2024openp5}, random indexing assigns item ids randomly and due to the way tokenizers used by LLMs, these ids might be split into sub-chunks. For example, item ids "2048" and "2049" could be split into ["20", "48"] and ["20", "49"], leading to artificial similarities between unrelated items.
When analyzing sequential indexing, we observe a performance decline across all datasets for both Spanish and Turkish prompts, with the decline being more pronounced for Turkish prompts. 
For example, on the LastFM dataset for sequential recommendations, there is an approximately 15\% drop for Spanish prompts compared to a 29\% drop for Turkish prompts. This difference is expected due to the distinct linguistic characteristics of Turkish compared to English.
Overall, the tables demonstrate that using non-English prompts in an already trained LLM-based recommender system affects the recommendation performance negatively, especially for languages which have different characteristics than English.

\begin{table*}
\caption{Performance of the retrained model on LastFM dataset with Sequential (S) indexing, [seen (s), unseen (u)] prompts for [Sequential, Straightforward] recommendation in [EN: English, ES: Spanish, TR: Turkish] (H: Hitrate, N: NDCG) }\label{table:compare_retrained_prompts_lastfm}
\begin{center}
\begin{tabular}{c|c|c|cccc|cccc} 
 \toprule
\multirow{2}{*}{\parbox{3.5em}{\centering Indexing\\(Prompt)}} & 
\multirow{2}{*}{\parbox{3.5em}{\centering Model\\Type}} & 
 \multirow{2}{*}{Lang} & 
        \multicolumn{4}{c}{Sequential} & \multicolumn{4}{c}{Straightforward} \\ 
        \cline{4-7}  \cline{8-11} 
{} & {} & {}     & H@5 & N@5 & H@10 & N@10 &  H@5 & N@5 & H@10 & N@10 \\
\midrule 
\multirow{6}{*}{S (s)} & \multirow{3}{*}{Retrained} & EN & 0.0229 & 0.0155 & 0.0413 & 0.0214  & 0.0376 & 0.0234 & 0.0523 & 0.0281\\
{}& {}  & ES & 0.0183 & 0.0136 & 0.0413 & 0.0211 & 0.0404 & 0.0281 & 0.0523 & 0.0320\\
{}& {}  & TR & 0.0257 & 0.0163 & 0.0459 & 0.0226 & 0.0367 & 0.0243 & 0.0523 & 0.0293 \\

\noalign{\vspace{0.3ex}}
\cline{2-11} 
\noalign{\vspace{0.3ex}}

{} & \multirow{3}{*}{Original} & EN & 0.0395 & 0.0262 & 0.0587 & 0.0323  & 0.0376 & 0.0259 & 0.0661 & 0.0350 \\
{}& {} & ES & 0.0321 & 0.0215 & 0.0523 & 0.0280 & 0.0330 & 0.0212 & 0.0578 & 0.0293 \\
{}& {} & TR & 0.0275 & 0.0167 & 0.0459 & 0.0226 & 0.0312 & 0.0221 & 0.0523 & 0.0289 \\

\midrule
\midrule

\multirow{6}{*}{S (u)}  & \multirow{3}{*}{Retrained} & EN & 0.0248 & 0.0155 & 0.0431 & 0.0213  & 0.0385 & 0.0244 & 0.0523 & 0.0288\\
{} & {} & ES & 0.0266 & 0.0169 & 0.0431 & 0.0221 & 0.0376 & 0.0261 & 0.0505 & 0.0302 \\
{} & {} & TR & 0.0367 & 0.0260 & 0.0514 & 0.0309 & 0.0340 & 0.0229 & 0.0523 & 0.0288 \\

\noalign{\vspace{0.3ex}}
\cline{2-11} 
\noalign{\vspace{0.3ex}}

{} & \multirow{3}{*}{Original} & EN & 0.0403 & 0.0265 & 0.0606 & 0.0331 & 0.0403 & 0.0282 & 0.0679 & 0.0370 \\
{} & {}& ES & 0.0358 & 0.0230 & 0.0532 & 0.0288 & 0.0376 & 0.0239 & 0.0551 & 0.0297 \\
{} & {} & TR & 0.0275 & 0.0195 & 0.0468 & 0.0256 & 0.0330 & 0.0212 & 0.0505 & 0.0269 \\

\bottomrule
 \end{tabular}
\end{center}
\end{table*}

\subsection{Impact of English and Non-English Prompts on LLM Recommender Fine-Tuning}


In OpenP5, the model was fine-tuned with English prompts to create OpenP5-T5 model. In this section, we fine-tune the same model with English, Spanish, and Turkish prompts and compare its performance across these languages.
For the analysis, LastFM dataset with sequential indexing is used. The training is ran for ten epochs with the setup described in the experimental setup section. The performance comparison is presented in Table~\ref{table:compare_retrained_prompts_lastfm}. In the table, the performance from the original OpenP5-T5 model and retrained model with multi-language prompts are presented.

\paragraph{Observations}
When we evaluate the performance of the retrained model, we observe a decline in performance on English prompts. For instance, the Hitrate@10 metric drops from 0.0587 to 0.0413 when the prompts are seen and the goal is to make a sequential recommendation. Similarly, it decreases from 0.0679 to 0.0523 when the prompts are unseen and the goal is a straightforward recommendation.
However, with the retrained model, the performance across all languages becomes closer. For example, the Hitrate@10 values for English, Spanish, and Turkish prompts are 0.0413, 0.0413, and 0.0459, respectively, when the prompts are seen and the goal is sequential recommendation. For unseen prompts aimed at straightforward recommendation, the Hitrate@10 values are 0.0523, 0.0505, and 0.0523 for English, Spanish, and Turkish prompts, respectively.
Overall, the table demonstrates that retraining the model with prompts from different languages affects its performance across these languages.

\section{Analysis and Discussion} \label{discuss}

Large language models (LLMs) have become the dominant tools for various natural language processing tasks, and their integration into recommender systems is a growing trend. This integration empowers users to leverage the capabilities of LLMs in everyday interactions. However, despite efforts to create multilingual LLMs, these models are predominantly trained on English texts. This bias leads to better performance in high-resource languages like English, while lower-resource languages face significant challenges.
As LLM-based recommenders evolve, it is crucial to consider these linguistic disparities, alongside other known issues such as position bias, popularity bias, and hallucinations.

Traditionally, recommender systems often function as black boxes for end-users, who simply request recommendations and receive items without insight into the underlying processes.
However, we anticipate that these systems will evolve enabling users to interact through natural language in addition to traditional methods like clicks and page-scrolls. We propose four phases in this evolution, as illustrated in Figure \ref{fig:phases}: 
(i) Initial Phase: Users interact with recommender via clicks and scrolls, while the recommender system relies on traditional black-box models.
(ii) LLM Integration: LLMs are integrated into these systems, using prompts that are independent of the user's language. As the result, the prompts can be in one of the high-resource languages, such as English.
(iii) Prompt Template Interaction: Users begin to interact with LLM-based recommenders by filling out prompt templates in their native languages, alongside traditional interactions.
(iv) Natural Language Interaction: Users engage with recommenders through natural language requests alongside traditional interactions.  

In light of these considerations, this paper explores the impact of non-English prompts on the performance of LLM-based recommendation systems in two folds: (i) by using non-English prompts on an already trained LLM-based recommender, and (ii) by incorporating both English and non-English prompts during further training. We selected Spanish and Turkish, which have distinct linguistic characteristics and resourcedness levels, for our experiments. 
The results revealed that non-English prompts negatively affect performance, especially in languages less similar to English. However, retraining the model with prompts from multiple languages led to a more balanced performance across all tested languages, despite a slight decrease in performance for English prompts.

\section{Conclusion}\label{conclusion}

Recommender systems suggest items to users based on their preferences, employing techniques from traditional methods like collaborative and content-based filtering to advanced deep learning approaches. Recently, large language models (LLMs) have been utilized to enhance recommendations through their generative capabilities, offering more personalized suggestions. Most state-of-the-art LLMs are designed for high-resource languages like English, which are well-studied and have abundant datasets.
In this work, we investigate the impact of non-English prompts on LLM-based recommenders by (i) applying non-English prompts to an already trained model, and (ii) retraining the model with both English and non-English prompts. Our findings reveal that non-English prompts can reduce performance on an existing LLM-based recommender, particularly for languages that have different characteristics than English. However, retraining the model with these languages results in more balanced performance across them, though with a slight decrease in performance for English prompts.

This preliminary work explores how prompt language affects LLM-based recommender systems. We tested English, Spanish, and Turkish prompts, and plan to investigate additional languages with varying resource levels and characteristics in future research. Developing evaluation sets in these languages could aid other researchers and improve recommendations for users in their native languages, which we aim to address in future work. As newer multilingual models emerge, future research should consider these models for recommendation tasks and evaluate their performance on low-resource languages.


\bibliography{refs}

\begin{thebibliography}{57}
\expandafter\ifx\csname natexlab\endcsname\relax\def\natexlab#1{#1}\fi
\providecommand{\url}[1]{\texttt{#1}}
\providecommand{\href}[2]{#2}
\providecommand{\path}[1]{#1}
\providecommand{\DOIprefix}{doi:}
\providecommand{\ArXivprefix}{arXiv:}
\providecommand{\URLprefix}{URL: }
\providecommand{\Pubmedprefix}{pmid:}
\providecommand{\doi}[1]{\href{http://dx.doi.org/#1}{\path{#1}}}
\providecommand{\Pubmed}[1]{\href{pmid:#1}{\path{#1}}}
\providecommand{\bibinfo}[2]{#2}
\ifx\xfnm\relax \def\xfnm[#1]{\unskip,\space#1}\fi
\bibitem[{Pan et~al.(2008)Pan, Zhou, Cao, Liu, Lukose, Scholz, and Yang}]{pan2008one}
\bibinfo{author}{R.~Pan}, \bibinfo{author}{Y.~Zhou}, \bibinfo{author}{B.~Cao}, \bibinfo{author}{N.~N. Liu}, \bibinfo{author}{R.~Lukose}, \bibinfo{author}{M.~Scholz}, \bibinfo{author}{Q.~Yang},
\newblock \bibinfo{title}{One-class collaborative filtering},
\newblock in: \bibinfo{booktitle}{2008 Eighth IEEE international conference on data mining}, \bibinfo{organization}{IEEE}, \bibinfo{year}{2008}, pp. \bibinfo{pages}{502--511}.
\bibitem[{Ye et~al.(2010)Ye, Yin, and Lee}]{Ye2010}
\bibinfo{author}{M.~Ye}, \bibinfo{author}{P.~Yin}, \bibinfo{author}{W.-C. Lee},
\newblock \bibinfo{title}{Location recommendation for location-based social networks},
\newblock in: \bibinfo{booktitle}{Proceedings of the 18th SIGSPATIAL International Conference on Advances in Geographic Information Systems}, GIS '10, \bibinfo{publisher}{ACM}, \bibinfo{address}{New York, NY, USA}, \bibinfo{year}{2010}, pp. \bibinfo{pages}{458--461}.
\bibitem[{Li et~al.(2015)Li, Cong, Li, Pham, and Krishnaswamy}]{li2015rank}
\bibinfo{author}{X.~Li}, \bibinfo{author}{G.~Cong}, \bibinfo{author}{X.-L. Li}, \bibinfo{author}{T.-A.~N. Pham}, \bibinfo{author}{S.~Krishnaswamy},
\newblock \bibinfo{title}{Rank-geofm: A ranking based geographical factorization method for point of interest recommendation},
\newblock in: \bibinfo{booktitle}{Proceedings of the 38th International ACM SIGIR Conference on Research and Development in Information Retrieval}, \bibinfo{organization}{ACM}, \bibinfo{year}{2015}, pp. \bibinfo{pages}{433--442}.
\bibitem[{He et~al.(2016)He, Li, Liao, Song, and Cheung}]{HeLLSC16}
\bibinfo{author}{J.~He}, \bibinfo{author}{X.~Li}, \bibinfo{author}{L.~Liao}, \bibinfo{author}{D.~Song}, \bibinfo{author}{W.~K. Cheung},
\newblock \bibinfo{title}{Inferring a personalized next point-of-interest recommendation model with latent behavior patterns},
\newblock in: \bibinfo{booktitle}{Proceedings of the Thirtieth {AAAI} Conference on Artificial Intelligence, February 12-17, 2016, Phoenix, Arizona, {USA.}}, \bibinfo{year}{2016}, pp. \bibinfo{pages}{137--143}.
\bibitem[{Ozsoy(2016)}]{ozsoy2016word}
\bibinfo{author}{M.~G. Ozsoy},
\newblock \bibinfo{title}{From word embeddings to item recommendation},
\newblock \bibinfo{journal}{arXiv preprint arXiv:1601.01356}  (\bibinfo{year}{2016}).
\bibitem[{Vasile et~al.(2016)Vasile, Smirnova, and Conneau}]{vasile2016meta}
\bibinfo{author}{F.~Vasile}, \bibinfo{author}{E.~Smirnova}, \bibinfo{author}{A.~Conneau},
\newblock \bibinfo{title}{Meta-prod2vec: Product embeddings using side-information for recommendation},
\newblock in: \bibinfo{booktitle}{Proceedings of the 10th ACM conference on recommender systems}, \bibinfo{year}{2016}, pp. \bibinfo{pages}{225--232}.
\bibitem[{He et~al.(2017)He, Liao, Zhang, Nie, Hu, and Chua}]{he2017neural}
\bibinfo{author}{X.~He}, \bibinfo{author}{L.~Liao}, \bibinfo{author}{H.~Zhang}, \bibinfo{author}{L.~Nie}, \bibinfo{author}{X.~Hu}, \bibinfo{author}{T.-S. Chua},
\newblock \bibinfo{title}{Neural collaborative filtering},
\newblock in: \bibinfo{booktitle}{Proceedings of the 26th international conference on world wide web}, \bibinfo{year}{2017}, pp. \bibinfo{pages}{173--182}.
\bibitem[{Musto et~al.(2018)Musto, Franza, Semeraro, de~Gemmis, and Lops}]{musto2018deep}
\bibinfo{author}{C.~Musto}, \bibinfo{author}{T.~Franza}, \bibinfo{author}{G.~Semeraro}, \bibinfo{author}{M.~de~Gemmis}, \bibinfo{author}{P.~Lops},
\newblock \bibinfo{title}{Deep content-based recommender systems exploiting recurrent neural networks and linked open data},
\newblock in: \bibinfo{booktitle}{Adjunct Publication of the 26th Conference on User Modeling, Adaptation and Personalization}, \bibinfo{organization}{ACM}, \bibinfo{year}{2018}, pp. \bibinfo{pages}{239--244}.
\bibitem[{Zheng et~al.(2018)Zheng, Zhang, Zheng, Xiang, Yuan, Xie, and Li}]{zheng2018drn}
\bibinfo{author}{G.~Zheng}, \bibinfo{author}{F.~Zhang}, \bibinfo{author}{Z.~Zheng}, \bibinfo{author}{Y.~Xiang}, \bibinfo{author}{N.~J. Yuan}, \bibinfo{author}{X.~Xie}, \bibinfo{author}{Z.~Li},
\newblock \bibinfo{title}{Drn: A deep reinforcement learning framework for news recommendation},
\newblock in: \bibinfo{booktitle}{Proceedings of the 2018 world wide web conference}, \bibinfo{year}{2018}, pp. \bibinfo{pages}{167--176}.
\bibitem[{Shi et~al.(2021)Shi, Tragos, Ozsoy, Dong, Hurley, Smyth, and Lawlor}]{shi2021dares}
\bibinfo{author}{B.~Shi}, \bibinfo{author}{E.~Z. Tragos}, \bibinfo{author}{M.~G. Ozsoy}, \bibinfo{author}{R.~Dong}, \bibinfo{author}{N.~Hurley}, \bibinfo{author}{B.~Smyth}, \bibinfo{author}{A.~Lawlor},
\newblock \bibinfo{title}{Dares: an asynchronous distributed recommender system using deep reinforcement learning},
\newblock \bibinfo{journal}{IEEE access} \bibinfo{volume}{9} (\bibinfo{year}{2021}) \bibinfo{pages}{83340--83354}.
\bibitem[{Sun et~al.(2019)Sun, Liu, Wu, Pei, Lin, Ou, and Jiang}]{sun2019bert4rec}
\bibinfo{author}{F.~Sun}, \bibinfo{author}{J.~Liu}, \bibinfo{author}{J.~Wu}, \bibinfo{author}{C.~Pei}, \bibinfo{author}{X.~Lin}, \bibinfo{author}{W.~Ou}, \bibinfo{author}{P.~Jiang},
\newblock \bibinfo{title}{Bert4rec: Sequential recommendation with bidirectional encoder representations from transformer},
\newblock in: \bibinfo{booktitle}{Proceedings of the 28th ACM international conference on information and knowledge management}, \bibinfo{year}{2019}, pp. \bibinfo{pages}{1441--1450}.
\bibitem[{Fan et~al.(2023)Fan, Zhao, Li, Liu, Mei, Wang, Tang, and Li}]{fan2023recommender}
\bibinfo{author}{W.~Fan}, \bibinfo{author}{Z.~Zhao}, \bibinfo{author}{J.~Li}, \bibinfo{author}{Y.~Liu}, \bibinfo{author}{X.~Mei}, \bibinfo{author}{Y.~Wang}, \bibinfo{author}{J.~Tang}, \bibinfo{author}{Q.~Li},
\newblock \bibinfo{title}{Recommender systems in the era of large language models (llms)},
\newblock \bibinfo{journal}{arXiv preprint arXiv:2307.02046}  (\bibinfo{year}{2023}).
\bibitem[{Geng et~al.(2022)Geng, Liu, Fu, Ge, and Zhang}]{geng2022recommendation}
\bibinfo{author}{S.~Geng}, \bibinfo{author}{S.~Liu}, \bibinfo{author}{Z.~Fu}, \bibinfo{author}{Y.~Ge}, \bibinfo{author}{Y.~Zhang},
\newblock \bibinfo{title}{Recommendation as language processing (rlp): A unified pretrain, personalized prompt \& predict paradigm (p5)},
\newblock in: \bibinfo{booktitle}{Proceedings of the 16th ACM Conference on Recommender Systems}, \bibinfo{year}{2022}, pp. \bibinfo{pages}{299--315}.
\bibitem[{Xu et~al.(2024)Xu, Hua, and Zhang}]{xu2024openp5}
\bibinfo{author}{S.~Xu}, \bibinfo{author}{W.~Hua}, \bibinfo{author}{Y.~Zhang},
\newblock \bibinfo{title}{Openp5: An open-source platform for developing, training, and evaluating llm-based recommender systems},
\newblock in: \bibinfo{booktitle}{Proceedings of the 47th International ACM SIGIR Conference on Research and Development in Information Retrieval}, \bibinfo{year}{2024}, pp. \bibinfo{pages}{386--394}.
\bibitem[{Ngo and Nguyen(2024)}]{ngo2024recgpt}
\bibinfo{author}{H.~Ngo}, \bibinfo{author}{D.~Q. Nguyen},
\newblock \bibinfo{title}{Recgpt: Generative pre-training for text-based recommendation},
\newblock \bibinfo{journal}{arXiv preprint arXiv:2405.12715}  (\bibinfo{year}{2024}).
\bibitem[{Nicholas and Bhatia(2023)}]{nicholas2023lost}
\bibinfo{author}{G.~Nicholas}, \bibinfo{author}{A.~Bhatia},
\newblock \bibinfo{title}{Lost in translation: large language models in non-english content analysis},
\newblock \bibinfo{journal}{arXiv preprint arXiv:2306.07377}  (\bibinfo{year}{2023}).
\bibitem[{Zhao et~al.(2024)Zhao, Zhang, Zhang, Gui, and Huang}]{zhao2024llama}
\bibinfo{author}{J.~Zhao}, \bibinfo{author}{Z.~Zhang}, \bibinfo{author}{Q.~Zhang}, \bibinfo{author}{T.~Gui}, \bibinfo{author}{X.~Huang},
\newblock \bibinfo{title}{Llama beyond english: An empirical study on language capability transfer},
\newblock \bibinfo{journal}{arXiv preprint arXiv:2401.01055}  (\bibinfo{year}{2024}).
\bibitem[{team(2024)}]{Cohere24AILangGap}
\bibinfo{author}{C.~F.~A. team},
\newblock \bibinfo{title}{The ai language gap},
\newblock \bibinfo{journal}{Policy Primer}  (\bibinfo{year}{2024}). \bibinfo{note}{\url{https://cohere.com/research/papers/the-ai-language-gap.pdf}}.
\bibitem[{Xu et~al.(2023)Xu, Hua, and Zhang}]{xu2023openp5}
\bibinfo{author}{S.~Xu}, \bibinfo{author}{W.~Hua}, \bibinfo{author}{Y.~Zhang},
\newblock \bibinfo{title}{Openp5: Benchmarking foundation models for recommendation},
\newblock \bibinfo{journal}{arXiv preprint arXiv:2306.11134}  (\bibinfo{year}{2023}).
\bibitem[{Qiu et~al.(2021)Qiu, Wu, Gao, and Fan}]{qiu2021u}
\bibinfo{author}{Z.~Qiu}, \bibinfo{author}{X.~Wu}, \bibinfo{author}{J.~Gao}, \bibinfo{author}{W.~Fan},
\newblock \bibinfo{title}{U-bert: Pre-training user representations for improved recommendation},
\newblock in: \bibinfo{booktitle}{Proceedings of the AAAI Conference on Artificial Intelligence}, volume~\bibinfo{volume}{35}, \bibinfo{year}{2021}, pp. \bibinfo{pages}{4320--4327}.
\bibitem[{Zhang et~al.(2022)Zhang, Zheng, and Wang}]{zhang2022gbert}
\bibinfo{author}{S.~Zhang}, \bibinfo{author}{N.~Zheng}, \bibinfo{author}{D.~Wang},
\newblock \bibinfo{title}{Gbert: Pre-training user representations for ephemeral group recommendation},
\newblock in: \bibinfo{booktitle}{Proceedings of the 31st ACM International Conference on Information \& Knowledge Management}, \bibinfo{year}{2022}, pp. \bibinfo{pages}{2631--2639}.
\bibitem[{Li et~al.(2023)Li, Wang, Li, Fu, Shen, Shang, and McAuley}]{li2023text}
\bibinfo{author}{J.~Li}, \bibinfo{author}{M.~Wang}, \bibinfo{author}{J.~Li}, \bibinfo{author}{J.~Fu}, \bibinfo{author}{X.~Shen}, \bibinfo{author}{J.~Shang}, \bibinfo{author}{J.~McAuley},
\newblock \bibinfo{title}{Text is all you need: Learning language representations for sequential recommendation},
\newblock in: \bibinfo{booktitle}{Proceedings of the 29th ACM SIGKDD Conference on Knowledge Discovery and Data Mining}, \bibinfo{year}{2023}, pp. \bibinfo{pages}{1258--1267}.
\bibitem[{Li et~al.(2024)Li, Zhang, Liu, and Chen}]{li2024large}
\bibinfo{author}{L.~Li}, \bibinfo{author}{Y.~Zhang}, \bibinfo{author}{D.~Liu}, \bibinfo{author}{L.~Chen},
\newblock \bibinfo{title}{Large language models for generative recommendation: A survey and visionary discussions},
\newblock in: \bibinfo{booktitle}{Proceedings of the 2024 Joint International Conference on Computational Linguistics, Language Resources and Evaluation (LREC-COLING 2024)}, \bibinfo{year}{2024}, pp. \bibinfo{pages}{10146--10159}.
\bibitem[{Vats et~al.(2024)Vats, Jain, Raja, and Chadha}]{vats2024exploring}
\bibinfo{author}{A.~Vats}, \bibinfo{author}{V.~Jain}, \bibinfo{author}{R.~Raja}, \bibinfo{author}{A.~Chadha},
\newblock \bibinfo{title}{Exploring the impact of large language models on recommender systems: An extensive review},
\newblock \bibinfo{journal}{arXiv preprint arXiv:2402.18590}  (\bibinfo{year}{2024}).
\bibitem[{Dai et~al.(2023)Dai, Shao, Zhao, Yu, Si, Xu, Sun, Zhang, and Xu}]{dai2023uncovering}
\bibinfo{author}{S.~Dai}, \bibinfo{author}{N.~Shao}, \bibinfo{author}{H.~Zhao}, \bibinfo{author}{W.~Yu}, \bibinfo{author}{Z.~Si}, \bibinfo{author}{C.~Xu}, \bibinfo{author}{Z.~Sun}, \bibinfo{author}{X.~Zhang}, \bibinfo{author}{J.~Xu},
\newblock \bibinfo{title}{Uncovering chatgpt’s capabilities in recommender systems},
\newblock in: \bibinfo{booktitle}{Proceedings of the 17th ACM Conference on Recommender Systems}, \bibinfo{year}{2023}, pp. \bibinfo{pages}{1126--1132}.
\bibitem[{Sun et~al.(2023)Sun, Yan, Ma, Wang, Ren, Chen, Yin, and Ren}]{sun2023chatgpt}
\bibinfo{author}{W.~Sun}, \bibinfo{author}{L.~Yan}, \bibinfo{author}{X.~Ma}, \bibinfo{author}{S.~Wang}, \bibinfo{author}{P.~Ren}, \bibinfo{author}{Z.~Chen}, \bibinfo{author}{D.~Yin}, \bibinfo{author}{Z.~Ren},
\newblock \bibinfo{title}{Is chatgpt good at search? investigating large language models as re-ranking agents},
\newblock \bibinfo{journal}{arXiv preprint arXiv:2304.09542}  (\bibinfo{year}{2023}).
\bibitem[{Gao et~al.(2023)Gao, Sheng, Xiang, Xiong, Wang, and Zhang}]{gao2023chat}
\bibinfo{author}{Y.~Gao}, \bibinfo{author}{T.~Sheng}, \bibinfo{author}{Y.~Xiang}, \bibinfo{author}{Y.~Xiong}, \bibinfo{author}{H.~Wang}, \bibinfo{author}{J.~Zhang},
\newblock \bibinfo{title}{Chat-rec: Towards interactive and explainable llms-augmented recommender system},
\newblock \bibinfo{journal}{arXiv preprint arXiv:2303.14524}  (\bibinfo{year}{2023}).
\bibitem[{He et~al.(2023)He, Xie, Jha, Steck, Liang, Feng, Majumder, Kallus, and McAuley}]{he2023large}
\bibinfo{author}{Z.~He}, \bibinfo{author}{Z.~Xie}, \bibinfo{author}{R.~Jha}, \bibinfo{author}{H.~Steck}, \bibinfo{author}{D.~Liang}, \bibinfo{author}{Y.~Feng}, \bibinfo{author}{B.~P. Majumder}, \bibinfo{author}{N.~Kallus}, \bibinfo{author}{J.~McAuley},
\newblock \bibinfo{title}{Large language models as zero-shot conversational recommenders},
\newblock in: \bibinfo{booktitle}{Proceedings of the 32nd ACM international conference on information and knowledge management}, \bibinfo{year}{2023}, pp. \bibinfo{pages}{720--730}.
\bibitem[{Zhang et~al.(2023)Zhang, Bao, Zhang, Wang, Feng, and He}]{zhang2023chatgpt}
\bibinfo{author}{J.~Zhang}, \bibinfo{author}{K.~Bao}, \bibinfo{author}{Y.~Zhang}, \bibinfo{author}{W.~Wang}, \bibinfo{author}{F.~Feng}, \bibinfo{author}{X.~He},
\newblock \bibinfo{title}{Is chatgpt fair for recommendation? evaluating fairness in large language model recommendation},
\newblock in: \bibinfo{booktitle}{Proceedings of the 17th ACM Conference on Recommender Systems}, \bibinfo{year}{2023}, pp. \bibinfo{pages}{993--999}.
\bibitem[{Wang and Lim(2023)}]{wang2023zero}
\bibinfo{author}{L.~Wang}, \bibinfo{author}{E.-P. Lim},
\newblock \bibinfo{title}{Zero-shot next-item recommendation using large pretrained language models},
\newblock \bibinfo{journal}{arXiv preprint arXiv:2304.03153}  (\bibinfo{year}{2023}).
\bibitem[{Du et~al.(2024)Du, Luo, Yan, Wang, Liu, Zhu, Song, and Zhang}]{du2024enhancing}
\bibinfo{author}{Y.~Du}, \bibinfo{author}{D.~Luo}, \bibinfo{author}{R.~Yan}, \bibinfo{author}{X.~Wang}, \bibinfo{author}{H.~Liu}, \bibinfo{author}{H.~Zhu}, \bibinfo{author}{Y.~Song}, \bibinfo{author}{J.~Zhang},
\newblock \bibinfo{title}{Enhancing job recommendation through llm-based generative adversarial networks},
\newblock in: \bibinfo{booktitle}{Proceedings of the AAAI Conference on Artificial Intelligence}, volume~\bibinfo{volume}{38}, \bibinfo{year}{2024}, pp. \bibinfo{pages}{8363--8371}.
\bibitem[{Liu et~al.(2024)Liu, Chen, Sakai, and Wu}]{liu2024once}
\bibinfo{author}{Q.~Liu}, \bibinfo{author}{N.~Chen}, \bibinfo{author}{T.~Sakai}, \bibinfo{author}{X.-M. Wu},
\newblock \bibinfo{title}{Once: Boosting content-based recommendation with both open-and closed-source large language models},
\newblock in: \bibinfo{booktitle}{Proceedings of the 17th ACM International Conference on Web Search and Data Mining}, \bibinfo{year}{2024}, pp. \bibinfo{pages}{452--461}.
\bibitem[{Hou et~al.(2024)Hou, Zhang, Lin, Lu, Xie, McAuley, and Zhao}]{hou2024large}
\bibinfo{author}{Y.~Hou}, \bibinfo{author}{J.~Zhang}, \bibinfo{author}{Z.~Lin}, \bibinfo{author}{H.~Lu}, \bibinfo{author}{R.~Xie}, \bibinfo{author}{J.~McAuley}, \bibinfo{author}{W.~X. Zhao},
\newblock \bibinfo{title}{Large language models are zero-shot rankers for recommender systems},
\newblock in: \bibinfo{booktitle}{European Conference on Information Retrieval}, \bibinfo{organization}{Springer}, \bibinfo{year}{2024}, pp. \bibinfo{pages}{364--381}.
\bibitem[{Liu et~al.(2023)Liu, Liu, Zhou, Lv, Zhou, and Zhang}]{liu2023chatgpt}
\bibinfo{author}{J.~Liu}, \bibinfo{author}{C.~Liu}, \bibinfo{author}{P.~Zhou}, \bibinfo{author}{R.~Lv}, \bibinfo{author}{K.~Zhou}, \bibinfo{author}{Y.~Zhang},
\newblock \bibinfo{title}{Is chatgpt a good recommender? a preliminary study},
\newblock \bibinfo{journal}{arXiv preprint arXiv:2304.10149}  (\bibinfo{year}{2023}).
\bibitem[{Tan et~al.(2024)Tan, Xu, Hua, Ge, Li, and Zhang}]{tan2024idgenrec}
\bibinfo{author}{J.~Tan}, \bibinfo{author}{S.~Xu}, \bibinfo{author}{W.~Hua}, \bibinfo{author}{Y.~Ge}, \bibinfo{author}{Z.~Li}, \bibinfo{author}{Y.~Zhang},
\newblock \bibinfo{title}{Idgenrec: Llm-recsys alignment with textual id learning},
\newblock in: \bibinfo{booktitle}{Proceedings of the 47th International ACM SIGIR Conference on Research and Development in Information Retrieval}, \bibinfo{year}{2024}, pp. \bibinfo{pages}{355--364}.
\bibitem[{Chu et~al.(2023)Chu, Hao, Ouyang, Wang, Wang, Shen, Gu, Cui, Li, Xue et~al.}]{chu2023leveraging}
\bibinfo{author}{Z.~Chu}, \bibinfo{author}{H.~Hao}, \bibinfo{author}{X.~Ouyang}, \bibinfo{author}{S.~Wang}, \bibinfo{author}{Y.~Wang}, \bibinfo{author}{Y.~Shen}, \bibinfo{author}{J.~Gu}, \bibinfo{author}{Q.~Cui}, \bibinfo{author}{L.~Li}, \bibinfo{author}{S.~Xue}, et~al.,
\newblock \bibinfo{title}{Leveraging large language models for pre-trained recommender systems},
\newblock \bibinfo{journal}{arXiv preprint arXiv:2308.10837}  (\bibinfo{year}{2023}).
\bibitem[{Bao et~al.(2023)Bao, Zhang, Zhang, Wang, Feng, and He}]{bao2023tallrec}
\bibinfo{author}{K.~Bao}, \bibinfo{author}{J.~Zhang}, \bibinfo{author}{Y.~Zhang}, \bibinfo{author}{W.~Wang}, \bibinfo{author}{F.~Feng}, \bibinfo{author}{X.~He},
\newblock \bibinfo{title}{Tallrec: An effective and efficient tuning framework to align large language model with recommendation},
\newblock in: \bibinfo{booktitle}{Proceedings of the 17th ACM Conference on Recommender Systems}, \bibinfo{year}{2023}, pp. \bibinfo{pages}{1007--1014}.
\bibitem[{Yang et~al.(2023)Yang, Chen, Jiang, Cho, Huang, and Lu}]{yang2023palr}
\bibinfo{author}{F.~Yang}, \bibinfo{author}{Z.~Chen}, \bibinfo{author}{Z.~Jiang}, \bibinfo{author}{E.~Cho}, \bibinfo{author}{X.~Huang}, \bibinfo{author}{Y.~Lu},
\newblock \bibinfo{title}{Palr: Personalization aware llms for recommendation},
\newblock \bibinfo{journal}{arXiv preprint arXiv:2305.07622}  (\bibinfo{year}{2023}).
\bibitem[{Li et~al.(2023)Li, Zhang, and Malthouse}]{li2023pbnr}
\bibinfo{author}{X.~Li}, \bibinfo{author}{Y.~Zhang}, \bibinfo{author}{E.~C. Malthouse},
\newblock \bibinfo{title}{Pbnr: Prompt-based news recommender system},
\newblock \bibinfo{journal}{arXiv preprint arXiv:2304.07862}  (\bibinfo{year}{2023}).
\bibitem[{Zhang et~al.(2023)Zhang, Xie, Hou, Zhao, Lin, and Wen}]{zhang2023recommendation}
\bibinfo{author}{J.~Zhang}, \bibinfo{author}{R.~Xie}, \bibinfo{author}{Y.~Hou}, \bibinfo{author}{W.~X. Zhao}, \bibinfo{author}{L.~Lin}, \bibinfo{author}{J.-R. Wen},
\newblock \bibinfo{title}{Recommendation as instruction following: A large language model empowered recommendation approach},
\newblock \bibinfo{journal}{arXiv preprint arXiv:2305.07001}  (\bibinfo{year}{2023}).
\bibitem[{Ji et~al.(2024)Ji, Li, Xu, Hua, Ge, Tan, and Zhang}]{ji2024genrec}
\bibinfo{author}{J.~Ji}, \bibinfo{author}{Z.~Li}, \bibinfo{author}{S.~Xu}, \bibinfo{author}{W.~Hua}, \bibinfo{author}{Y.~Ge}, \bibinfo{author}{J.~Tan}, \bibinfo{author}{Y.~Zhang},
\newblock \bibinfo{title}{Genrec: Large language model for generative recommendation},
\newblock in: \bibinfo{booktitle}{European Conference on Information Retrieval}, \bibinfo{organization}{Springer}, \bibinfo{year}{2024}, pp. \bibinfo{pages}{494--502}.
\bibitem[{Wang et~al.(2024)Wang, Du, Sun, Chua, Feng, Wang, and Zhang}]{wang2024re2llm}
\bibinfo{author}{Z.~Wang}, \bibinfo{author}{Y.~Du}, \bibinfo{author}{Z.~Sun}, \bibinfo{author}{H.~Chua}, \bibinfo{author}{K.~Feng}, \bibinfo{author}{W.~Wang}, \bibinfo{author}{J.~Zhang},
\newblock \bibinfo{title}{Re2llm: Reflective reinforcement large language model for session-based recommendation},
\newblock \bibinfo{journal}{arXiv preprint arXiv:2403.16427}  (\bibinfo{year}{2024}).
\bibitem[{Wu et~al.(2024{\natexlab{a}})Wu, Qiu, Zheng, Zhu, and Chen}]{wu2024exploring}
\bibinfo{author}{L.~Wu}, \bibinfo{author}{Z.~Qiu}, \bibinfo{author}{Z.~Zheng}, \bibinfo{author}{H.~Zhu}, \bibinfo{author}{E.~Chen},
\newblock \bibinfo{title}{Exploring large language model for graph data understanding in online job recommendations},
\newblock in: \bibinfo{booktitle}{Proceedings of the AAAI Conference on Artificial Intelligence}, volume~\bibinfo{volume}{38}, \bibinfo{year}{2024}{\natexlab{a}}, pp. \bibinfo{pages}{9178--9186}.
\bibitem[{Wu et~al.(2024{\natexlab{b}})Wu, Xie, Zhu, Zhuang, Zhang, Lin, and He}]{wu2024personalized}
\bibinfo{author}{Y.~Wu}, \bibinfo{author}{R.~Xie}, \bibinfo{author}{Y.~Zhu}, \bibinfo{author}{F.~Zhuang}, \bibinfo{author}{X.~Zhang}, \bibinfo{author}{L.~Lin}, \bibinfo{author}{Q.~He},
\newblock \bibinfo{title}{Personalized prompt for sequential recommendation},
\newblock \bibinfo{journal}{IEEE Transactions on Knowledge and Data Engineering}  (\bibinfo{year}{2024}{\natexlab{b}}).
\bibitem[{Wang et~al.(2022)Wang, Yu, Ma, Zhang, Chen, Liu, and Ma}]{wang2022towards}
\bibinfo{author}{C.~Wang}, \bibinfo{author}{Y.~Yu}, \bibinfo{author}{W.~Ma}, \bibinfo{author}{M.~Zhang}, \bibinfo{author}{C.~Chen}, \bibinfo{author}{Y.~Liu}, \bibinfo{author}{S.~Ma},
\newblock \bibinfo{title}{Towards representation alignment and uniformity in collaborative filtering},
\newblock in: \bibinfo{booktitle}{Proceedings of the 28th ACM SIGKDD conference on knowledge discovery and data mining}, \bibinfo{year}{2022}, pp. \bibinfo{pages}{1816--1825}.
\bibitem[{Wu et~al.(2023)Wu, Zheng, Qiu, Wang, Gu, Shen, Qin, Zhu, Zhu, Liu et~al.}]{wu2023survey}
\bibinfo{author}{L.~Wu}, \bibinfo{author}{Z.~Zheng}, \bibinfo{author}{Z.~Qiu}, \bibinfo{author}{H.~Wang}, \bibinfo{author}{H.~Gu}, \bibinfo{author}{T.~Shen}, \bibinfo{author}{C.~Qin}, \bibinfo{author}{C.~Zhu}, \bibinfo{author}{H.~Zhu}, \bibinfo{author}{Q.~Liu}, et~al.,
\newblock \bibinfo{title}{A survey on large language models for recommendation},
\newblock \bibinfo{journal}{arXiv preprint arXiv:2305.19860}  (\bibinfo{year}{2023}).
\bibitem[{Azamfirei et~al.(2023)Azamfirei, Kudchadkar, and Fackler}]{azamfirei2023large}
\bibinfo{author}{R.~Azamfirei}, \bibinfo{author}{S.~R. Kudchadkar}, \bibinfo{author}{J.~Fackler},
\newblock \bibinfo{title}{Large language models and the perils of their hallucinations},
\newblock \bibinfo{journal}{Critical Care} \bibinfo{volume}{27} (\bibinfo{year}{2023}) \bibinfo{pages}{120}.
\bibitem[{Tan et~al.(2023)Tan, Ge, Zhu, Xia, Luo, Ji, and Zhang}]{tan2023user}
\bibinfo{author}{J.~Tan}, \bibinfo{author}{Y.~Ge}, \bibinfo{author}{Y.~Zhu}, \bibinfo{author}{Y.~Xia}, \bibinfo{author}{J.~Luo}, \bibinfo{author}{J.~Ji}, \bibinfo{author}{Y.~Zhang},
\newblock \bibinfo{title}{User-controllable recommendation via counterfactual retrospective and prospective explanations},
\newblock in: \bibinfo{booktitle}{ECAI 2023}, \bibinfo{publisher}{IOS Press}, \bibinfo{year}{2023}, pp. \bibinfo{pages}{2307--2314}.
\bibitem[{Team(2023)}]{team2023internlm}
\bibinfo{author}{I.~Team}, \bibinfo{title}{Internlm: A multilingual language model with progressively enhanced capabilities}, \bibinfo{year}{2023}. \bibinfo{note}{\url{https://github.com/InternLM/InternLM}}.
\bibitem[{Cui et~al.(2023)Cui, Yang, and Yao}]{cui2023efficient}
\bibinfo{author}{Y.~Cui}, \bibinfo{author}{Z.~Yang}, \bibinfo{author}{X.~Yao},
\newblock \bibinfo{title}{Efficient and effective text encoding for chinese llama and alpaca},
\newblock \bibinfo{journal}{arXiv preprint arXiv:2304.08177}  (\bibinfo{year}{2023}).
\bibitem[{Acikgoz et~al.(2024)Acikgoz, Erdogan, and Yuret}]{acikgoz2024bridging}
\bibinfo{author}{E.~C. Acikgoz}, \bibinfo{author}{M.~Erdogan}, \bibinfo{author}{D.~Yuret},
\newblock \bibinfo{title}{Bridging the bosphorus: Advancing turkish large language models through strategies for low-resource language adaptation and benchmarking},
\newblock \bibinfo{journal}{arXiv preprint arXiv:2405.04685}  (\bibinfo{year}{2024}).
\bibitem[{Kesgin et~al.(2024)Kesgin, Yuce, Dogan, Uzun, Uz, Seyrek, Zeer, and Amasyali}]{kesgin2024introducing}
\bibinfo{author}{H.~T. Kesgin}, \bibinfo{author}{M.~K. Yuce}, \bibinfo{author}{E.~Dogan}, \bibinfo{author}{M.~E. Uzun}, \bibinfo{author}{A.~Uz}, \bibinfo{author}{H.~E. Seyrek}, \bibinfo{author}{A.~Zeer}, \bibinfo{author}{M.~F. Amasyali},
\newblock \bibinfo{title}{Introducing cosmosgpt: Monolingual training for turkish language models},
\newblock \bibinfo{journal}{arXiv preprint arXiv:2404.17336}  (\bibinfo{year}{2024}).
\bibitem[{Raffel et~al.(2020)Raffel, Shazeer, Roberts, Lee, Narang, Matena, Zhou, Li, and Liu}]{raffel2020exploring}
\bibinfo{author}{C.~Raffel}, \bibinfo{author}{N.~Shazeer}, \bibinfo{author}{A.~Roberts}, \bibinfo{author}{K.~Lee}, \bibinfo{author}{S.~Narang}, \bibinfo{author}{M.~Matena}, \bibinfo{author}{Y.~Zhou}, \bibinfo{author}{W.~Li}, \bibinfo{author}{P.~J. Liu},
\newblock \bibinfo{title}{Exploring the limits of transfer learning with a unified text-to-text transformer},
\newblock \bibinfo{journal}{Journal of machine learning research} \bibinfo{volume}{21} (\bibinfo{year}{2020}) \bibinfo{pages}{1--67}.
\bibitem[{Touvron et~al.(2023)Touvron, Martin, Stone, Albert, Almahairi, Babaei, Bashlykov, Batra, Bhargava, Bhosale et~al.}]{touvron2023llama}
\bibinfo{author}{H.~Touvron}, \bibinfo{author}{L.~Martin}, \bibinfo{author}{K.~Stone}, \bibinfo{author}{P.~Albert}, \bibinfo{author}{A.~Almahairi}, \bibinfo{author}{Y.~Babaei}, \bibinfo{author}{N.~Bashlykov}, \bibinfo{author}{S.~Batra}, \bibinfo{author}{P.~Bhargava}, \bibinfo{author}{S.~Bhosale}, et~al.,
\newblock \bibinfo{title}{Llama 2: Open foundation and fine-tuned chat models},
\newblock \bibinfo{journal}{arXiv preprint arXiv:2307.09288}  (\bibinfo{year}{2023}).
\bibitem[{Hu et~al.(2021)Hu, Shen, Wallis, Allen-Zhu, Li, Wang, Wang, and Chen}]{hu2021lora}
\bibinfo{author}{E.~J. Hu}, \bibinfo{author}{Y.~Shen}, \bibinfo{author}{P.~Wallis}, \bibinfo{author}{Z.~Allen-Zhu}, \bibinfo{author}{Y.~Li}, \bibinfo{author}{S.~Wang}, \bibinfo{author}{L.~Wang}, \bibinfo{author}{W.~Chen},
\newblock \bibinfo{title}{Lora: Low-rank adaptation of large language models},
\newblock \bibinfo{journal}{arXiv preprint arXiv:2106.09685}  (\bibinfo{year}{2021}).
\bibitem[{Joshi et~al.(2020)Joshi, Santy, Budhiraja, Bali, and Choudhury}]{joshi2020state}
\bibinfo{author}{P.~Joshi}, \bibinfo{author}{S.~Santy}, \bibinfo{author}{A.~Budhiraja}, \bibinfo{author}{K.~Bali}, \bibinfo{author}{M.~Choudhury},
\newblock \bibinfo{title}{The state and fate of linguistic diversity and inclusion in the nlp world},
\newblock \bibinfo{journal}{arXiv preprint arXiv:2004.09095}  (\bibinfo{year}{2020}).
\bibitem[{Dhamecha et~al.(2021)Dhamecha, Murthy~V, Bharadwaj, Sankaranarayanan, and Bhattacharyya}]{dhamecha2021role}
\bibinfo{author}{T.~I. Dhamecha}, \bibinfo{author}{R.~Murthy~V}, \bibinfo{author}{S.~Bharadwaj}, \bibinfo{author}{K.~Sankaranarayanan}, \bibinfo{author}{P.~Bhattacharyya},
\newblock \bibinfo{title}{Role of language relatedness in multilingual fine-tuning of language models: A case study in indo-aryan languages},
\newblock \bibinfo{journal}{arXiv preprint arXiv:2109.10534}  (\bibinfo{year}{2021}).

\end{thebibliography}




\end{document}